\documentclass[aps,prb,twocolumn,superscriptaddress,showpacs]{revtex4}
\usepackage{graphicx}
\usepackage{calc}
\usepackage{bm}

\begin{document}

\title{Spin ordering: two different scenarios for
the single and double layer structures in the fractional and
integer quantum Hall effect regimes}

\author{V.T.~Dolgopolov}
\email[Corresponding author. E-mail:~]{dolgop@issp.ac.ru}
\affiliation{Institute of Solid State Physics RAS, Chernogolovka,
Moscow District 142432, Russia}

\author{E.V.~Deviatov}
 \affiliation{Institute of Solid State
Physics RAS, Chernogolovka, Moscow District 142432, Russia}

\author{V.S.~Khrapai}
\affiliation{Institute of Solid State Physics RAS, Chernogolovka,
Moscow District 142432, Russia}
\affiliation{Ludwig-Maximilians-Universit\"{a}t M\"{u}nchen,
Geschwister-Scholl-Platz 1, D-80539 Munich Germany}

\author{D.~Reuter}
\affiliation{Lehrstuhl f\"ur Angewandte Festk\"orperphysik,
Ruhr-Universit\"at Bochum, Universit\"atsstrasse 150, D-44780
Bochum, Germany}

\author{A.D.~Wieck}
\affiliation{Lehrstuhl f\"ur Angewandte Festk\"orperphysik,
Ruhr-Universit\"at Bochum, Universit\"atsstrasse 150, D-44780
Bochum, Germany}

\author{A.~Wixforth}
\affiliation{Institut f\"ur Physik, Universitat Augsburg,
Universitatsstrasse, 1 D-86135 Augsburg,  Germany}

\author{K.L.~Campman}
\affiliation{Materials Department and Center for Quantized
Electronic Structures, University of California, Santa Barbara,
California 93106, USA}

\author{A.C.~Gossard}
\affiliation{Materials Department and Center for Quantized
Electronic Structures, University of California, Santa Barbara,
California 93106, USA}

\date{\today}

\begin{abstract}
We investigate the ground state competition at the transition from
the spin unpolarized to spin ordered phase at filling factor
$\nu=2/3$ in single layer heterostructure and at  $\nu=2$ in
double layer quantum well. To trace the quantum Hall phase we use
the minimum in the dissipative conductivity $\sigma_{xx}$. We
observe two different transition scenarios in two investigated
situations. For one of them we propose a qualitative explanation,
based on the domain structure evolution in the vicinity of the
transition point. The origin for the second scenario,
corresponding to the experimental situation at $\nu=2$ in double
layer 2DES, still remains unclear.
\end{abstract}


\maketitle

The change of the ground state at fixed filling factor was
experimentally observed in a number of two dimensional electron
systems (2DES) subjected to quantizing magnetic
fields~\cite{eisen,caf,smet,fisher}. There are two very prominent
examples of the ground state competition: (i) the phase transition
from the spin unpolarized into the canted antiferromagnetic phase
in double layer system~\cite{caf} at filling factor $\nu=2$ and
(ii) the transitions between spin unpolarised and fully spin
polarized states in the fractional quantum Hall effect (FQHE)
regime~\cite{smet}. Electron correlations (inter-plane and
in-plane, correspondingly) play a significant role in both cases,
allowing to combine them into a single class of physical
phenomena~\cite{macdonald}. The competition between ground states
still survives even at zero temperature, thus the above mentioned
transitions are caused by quantum fluctuations and are supposed to
be the quantum phase transitions~\cite{sachdev}.

Both the integer and the fractional quantum Hall effects are
caused by the gap in the 2DES electron spectrum and
disorder~\cite{prange}. Quantum Hall phase exists within the strip
in the $(B,n_s)$-plane, around the line of the corresponding
integer or fractional filling factor. Within the strip, the
electron density $n_s$ and the magnetic field $B$ can only affect
on the Fermi level position and have no radical influence on the
physical properties of the 2DES. In contrast to this situation, in
the vicinity of the phase transition point the ground state itself
is a function  of these two parameters $B,n_s$  and a complicated
behavior of the 2DES properties can be expected.

Under the FQHE conditions, in a strong perpendicular magnetic
field $B_{norm}$, corresponding to the particular fractional
filling factor $\nu=hcn_s/eB_{norm}$, the Hall conductivity is of
quantized value $\nu e^2/h$ while the longitudinal one vanishes in
the high quality 2DES. Electron spins can be considered as
parallel in the high-$B_{norm}$ limit, while at the simultaneous
lowering of $B_{norm}, n_s$ a transition into the partially spin
polarized or even spin unpolarized state is
predicted~\cite{chacra}. Qualitatively this transition can be
understood as a result of the competition between the Zeeman and
exchange energies in strongly correlated electron liquid.
Disappearance of the minimum in the dissipative conductivity
component at some electron density $n_s^{tr}$ and reappearance of
the minimum around $n_s^{tr}$ at constant fractional filling
factor was interpreted as the observation of the ground state
competition~\cite{eisen,smet}.

A very similar effect was theoretically predicted\cite{sarma} and
experimentally observed~\cite{caf,pellegr} in a double-layer
system with symmetric electron density distribution at integer
total filling factor  $\nu=2$. In this case, the competition of
different ground states is caused by the interplay between the
inter-plain Coulomb energy, the spin splitting, and the
symmetrical-antisymmetrical splitting. In the simplest
single-particle picture (disregarding the Coulomb interaction),
each Landau level has four sublevels, originating from the spin
and symmetrical-antisymmetrical splittings.  At total filling
factor $\nu=2$, increasing the spin splitting causes a transition
from the spin-unpolarized ground state, with anti-parallel spin
orientations of occupied sublevels, to the ferromagnetic one with
parallel spins. Near the transition point, the intralayer exchange
interaction mixes two lowest states of the electron system and
gives rise to the intermediate canted antiferromagnetic phase,
characterized by interlayer antiferromagnetic spin correlations.
It is easy to see the analogy to the spin transition in the FQHE
regime. Experimentally, both transitions can be forced, e.g.,  by
the parallel magnetic field component, which increases the Zeeman
energy and suppresses tunnelling and interlayer correlations in a
double layer system.

In the present paper we investigate a competition of the ground
states within the narrow strip in the ($B,n_s$)-plane near the
fixed filling factor $\nu$ in two different electron systems. They
are the single layer at fractional $\nu=2/3$ and the double
quantum well at integer $\nu=2$. We want to find common and
different features of the 2DES behavior in the vicinity of the
transition point.

Our samples are grown by molecular beam epitaxy on semi-insulating
GaAs substrate. Single-layer GaAs/AlGaAs heterostructure contains
a 2DEG located 150 nm below the surface. The mobility at 4K is
1.83 $\cdot 10^{6}$cm$^{2}$/Vs and the carrier density 8.49 $\cdot
10^{10}$cm$^{-2}$. Double-layer system is formed in a 760~\AA\
wide symmetrically doped parabolic quantum well, containing a
3-monolayer thick AlAs sheet grown in the center, which serves as
a tunnel barrier between both parts on either side. The
symmetric-antisymmetric splitting in the bilayer electron system
as determined from far infrared measurements and model
calculations~\cite{hart} is equal to $\Delta_{SAS}=1.3$~meV.
Samples were prepared from two different wafers (A and B) with
close growth parameters.

The samples were patterned in quasi-Corbino geometry~\cite{alida}
with the gate area about 0.5~mm$^2$. Ohmic contacts are made to
both  parts of the well in double-layer samples. We trace the
dissipative conductivity minimum near the fractional $\nu=2/3$ for
single-layer samples and near the integer $\nu=2$ for double-layer
ones in the $(n_s,B)$-plane by usual magnetoresistance and
magnetocapasitance measurements. The experiment is performed at
the temperature of 30~mK for different tilt angles of the magnetic
field with respect to normal to the interface.

\begin{figure}
\includegraphics[width= 0.75\columnwidth]{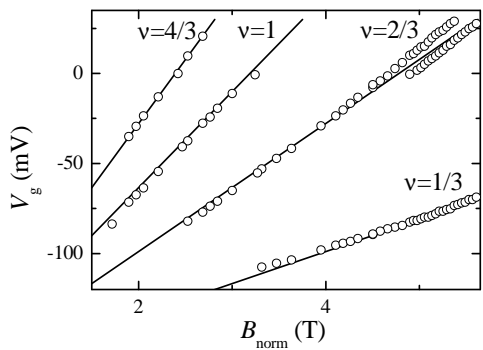}%
\caption{ Positions of the dissipative conductivity minima (open
circles) as function of the gate voltage $V_g$ (which controls the
electron concentration) and the normal magnetic field component
$B_{norm}$. Solid lines show the exact positions of integer and
fractional filling factors $\nu$. Magnetic field is tilted in
respect to the normal to the sample plane by the angle
$\alpha=19^\circ$. The phase transition at $\nu=2/3$ takes place
at $n_s^{tr}=8.77\cdot 10^{10}$~cm$^{-2}$ \label{fan19}}
\end{figure}

An example of the fan chart in ($n_s,B$)-plane for single layer
structure is shown in Fig.~\ref{fan19}. The fan chart lines for
integer and fractional quantum Hall states at $\nu=4/3;1;1/3$ do
not show any peculiarities. In contrast, the ground state at
$\nu=2/3$ is changing at the electron density $n_s^{tr}=8.77\cdot
10^{10}$~cm$^{-2}$. The comparison of this value with the known
from the previous experiments~\cite{eisen,smet} demonstrates that
electron density at the transition point $n_s^{tr}$ is sample
dependent. It is not surprising, because the electron-electron
interaction depends on the wave function extension in the
direction normal to the interface, which varies from sample to
sample. As expected, the transition point shifts towards the lower
electron density while increasing the parallel to the interface
field component $B_{par}$. From our experimental data the
derivative $d n_s^{tr}/d B_{par}$ can be estimated as $10^{10}$
(cm$^2$ T)$^{-1}$.

\begin{figure}
\includegraphics[width= 0.75\columnwidth]{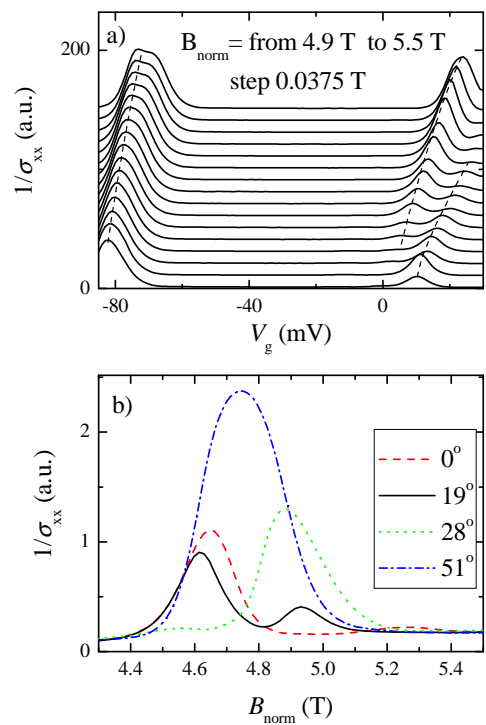}%
\caption{ (a) $1 / \sigma_{xx}$ as function of the gate voltage
$V_g$ (electron concentration) at different magnetic fields
$B_{norm}$. Magnetic field is tilted in respect to the normal to
the sample plane by the angle $\alpha=19^\circ$. Curves are
shifted for clarity. Dash highlights the minima positions (b) $1 /
\sigma_{xx}$ as function of the perpendicular to the sample plane
magnetic field component $B_{norm}$ at different tilt angles
$\alpha$: $0^\circ$(dash), $19^\circ$ (solid), $28^\circ$ (dots),
$51^\circ$ (dash-dot). \label{reconstr}}
\end{figure}

In the vicinity of the transition point two minima in
$\sigma_{xx}$ are observable, see Fig.~\ref{reconstr}. One of
them, corresponding to the upper branch in Fig.~\ref{fan19}, is
the continuation of the $\nu=2/3$ line at low electron density,
the second is connected with this $\nu=2/3$ line at high density.
In some region near the transition point $n_s^{tr}$ two minima in
the dissipative conductivity can be found in $\sigma_{xx}(n_s)$
sweep at fixed $B$ (see Fig.~\ref{reconstr} a) or on in
$\sigma_{xx}(B)$ sweep at fixed $n_s$ (see Fig.~\ref{reconstr} b).

\begin{figure}
\includegraphics[width= 0.75\columnwidth]{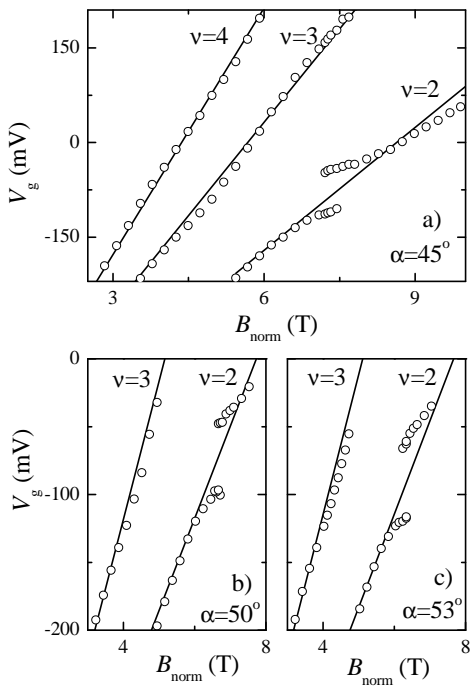}%
\caption{ Positions of the dissipative conductivity minima (open
circles) as function of gate voltage $V_g$ (which controls the
electron concentration) and perpendicular magnetic field component
$B_{norm}$ for two different wafers A (a) and B (b,c). Solid lines
show the exact positions of integer filling factors $\nu$. The
tilt angle of magnetic field with respect to normal to the sample
plane equals $\alpha=45^\circ$ (a) and $\alpha=50^\circ,53^\circ$
(b,c). The phase transition at $\nu=2$ takes place at
$n_s^{tr}=3.63\cdot 10^{11}$ cm$^{-2}$\label{vadik}}
\end{figure}

At the first glance, the behavior of the double-layer system is
very similar, see Fig.~\ref{vadik}. The phase transition is
observed in tilted magnetic fields for samples from wafers A
(Fig.~\ref{vadik} a) and B (Fig.~\ref{vadik} b),c) . Under the
same conditions no peculiarities are found at filling factors
$\nu=3,4$. At tilt angles $45^\circ,50^\circ,53^\circ$ the ground
state for $\nu=2$ is changing at the electron density
$n_s^{tr}=3.63\cdot 10^{11}$ cm$^{-2}$. The density $n_s^{tr}$ is
also sample dependent, nevertheless, qualitatively all
observations are sample independent, as it is easy to see from
comparison  of Figs.~\ref{vadik} a),b),c). The derivative $d
n_s^{tr}/d B_{par} \sim 4\cdot 10^{10}$ (cm$^2$ T)$^{-1}$. The
same order  of the value $d n_s^{tr}/d B_{par}$ as in the FQHE
case means that we deal with similar competition between Coulomb
and Zeeman energy in both cases.

Remarkably, the symmetry of $\sigma_{xx}$-minima positions in
Fig.~\ref{fan19} and Fig.~\ref{vadik} is totally different. In
Fig.~\ref{vadik} the upper branch is the continuation of the
$\nu=2$ line at high electron density and the bottom one  is
connected with this line at low density. Such a scenario is
totally different from the FQHE case.

We have to mention that the existence of two minima in the
($B,n_s$)-plane is non-trivial and needs in explanation. We
propose here the explanation, based on the consideration of domain
structure in the vicinity of the transition point $n_s^{tr}$. It
is well known and clearly shown experimentally, that in both 
cases the complicated domain structure does exist in the vicinity
of the transition point. The area covered by domains of one phase
is a function of the filling factor $\nu$. At
 the points $n_s^{tr}$, the areas
covered by different phases are equal and the system demonstrates
non-zero dissipative conductivity due to the percolation in phase
boundaries. One can expect the appearance of the deep minimum in
$\sigma_{xx}$ if the domains, belonging to one of the phase, would
create an infinite cluster. Because the Zeeman splitting is
smaller in weak magnetic fields, it is natural to expect that
domains with low-field configuration prevails at filling factors
above $\nu=2/3$ and $\nu=2$. Such the way of explanation seems to
give an adequate description for the diagrams corresponding to
filling factor $\nu=2/3$ in the single layer 2DES, but hardly can
describe the observation at $\nu=2$ in double layer systems.

In conclusion, we investigate the ground state competition at the
transition from the spin unpolarized to spin ordered phase at
filing factor  $\nu=2/3$ in single layer heterostructure and at
$\nu=2$ in double layer quantum well. To trace the quantum Hall
effect phase we use the minimum in the dissipative conductivity
$\sigma_{xx}$. We observe two different transition scenarios in
two investigated situations. For one of them we propose a
qualitative explanation, based on the domain structure evolution
in the vicinity of the transition point. The origin for the second
scenario, corresponding to the experimental situation at $\nu=2$
in double layer 2DES, still remains unclear.

We wish to thank A.A.~Shashkin and A.A.~Kapustin for help during
the experiment. We gratefully acknowledge financial support by the
RFBR, RAS, the Programme "The State Support of Leading Scientific
Schools". E.V.D. acknowledges support by Russian Science Support
Foundation. V.S.K. thanks Alexander von Humboldt Stiftung for a
financial support.

\end{document}